\documentclass[10pt]{article}

\usepackage{amsmath,amsmath,amssymb}
\usepackage{mathrsfs}
\usepackage{multirow}
\usepackage{graphicx}
\usepackage{authblk}
\usepackage{indentfirst}
\usepackage{multicol}  
\usepackage{tabu}
\usepackage{tabularx}
\usepackage{url}
\usepackage{fancyhdr}
\usepackage[numbers]{natbib}
\usepackage{lineno}
\usepackage{makecell}
\usepackage{fancyhdr}
\usepackage{colortbl}
\usepackage{caption}

\usepackage{hyperref}
\hypersetup{colorlinks=true,linkcolor=blue,anchorcolor=blue,citecolor=blue}

\title{Superionic Ionic Conductor Discovery via Multiscale Topological Learning}

\author[1,2]{Dong Chen\thanks{contributed equally to this work}} 
\author[1]{Bingxu Wang\thanks{contributed equally to this work}} 
\author[1]{Shunning Li} 
\author[1]{Wentao Zhang} 
\author[1]{Kai Yang} 
\author[1]{Yongli Song} 
\author[2,3,4]{Guo-Wei Wei \thanks{Corresponding author: weig@msu.edu}}
\author[1]{Feng Pan \thanks{Corresponding author: panfeng@pkusz.edu.cn}}
\affil[1]{School of Advanced Materials, Peking University, Shenzhen Graduate School, Shenzhen 518055, China}
\affil[2]{Department of Mathematics, Michigan State University, MI, 48824, USA}
\affil[3]{Department of Electrical and Computer Engineering, Michigan State University, MI 48824, USA}
\affil[4]{Department of Biochemistry and Molecular Biology, Michigan State University, MI 48824, USA}

\date{}

\begin{document}
    \maketitle

    \paragraph{Abstract} 

        Lithium superionic conductors (LSICs) are crucial for next-generation solid-state batteries, offering exceptional ionic conductivity and enhanced safety for renewable energy and electric vehicles. However, their discovery is extremely challenging due to the vast chemical space, limited labeled data, and the understanding of complex structure-function relationships required for optimizing ion transport. This study introduces a multiscale topological learning (MTL) framework, integrating algebraic topology and unsupervised learning to tackle these challenges efficiently. By modeling lithium-only and lithium-free substructures, the framework extracts multiscale topological features and introduces two topological screening metrics-cycle density and minimum connectivity distance-to ensure structural connectivity and ion diffusion compatibility. Promising candidates are clustered via unsupervised algorithms to identify those resembling known superionic conductors. For final refinement, candidates that pass chemical screening undergo ab initio molecular dynamics simulations for validation. This approach led to the discovery of 14 novel LSICs, four of which have been independently validated in recent experiments. This success accelerates the identification of LSICs and demonstrates broad adaptability, offering a scalable tool for addressing complex materials discovery challenges.

  
    \newpage
	
    \section{Introduction}


    The discovery of superionic conductors—materials with exceptional ion transport properties—is crucial for advancing electrochemical energy storage and conversion technologies, including batteries\cite{dunn2011electrical, seino2014sulphide, thangadurai2014garnet}, fuel cells\cite{wachsman2011lowering}, and ceramic membranes\cite{xu2005ion, randau2020benchmarking}. Among these, lithium superionic conductors (LSICs) are particularly promising alternatives to conventional organic liquid electrolytes due to their high ionic conductivity, broad electrochemical stability, and enhanced safety\cite{janek2016solid}. These attributes are vital for improving the performance, energy density, and lifespan of lithium-ion batteries. However, the discovery of LSICs remains a significant challenge. Only a limited number of lithium-based compounds, such as Li$_{10}$GeP$_{2}$S$_{12}$ (LGPS)\cite{zhao2016new}, garnet Li$_{7}$La$_{3}$Zr$_{2}$O$_{12}$ (LLZO)\cite{thangadurai2003novel,murugan2007fast}, NASICON\cite{arbi2007lithium}, and Li$_{1.3}$Al$_{0.3}$Ti$_{1.7}$(PO$_{4}$)$_{3}$ (LATP)\cite{kamaya2011lithium, aono1990ionic1}, exhibit room-temperature ionic conductivity comparable to liquid electrolytes. This limited number, coupled with insufficient ionic conductivity data, complicates the discovery of new LSICs. Furthermore, the experimental process to validate these materials is both expensive and time-consuming, and traditional computational methods, such as Density Functional Theory (DFT) and ab initio molecular dynamics (AIMD) simulations, are extremely expensive and intractable for large-scale screening. Despite their potential, current LSICs do not meet the comprehensive requirements for widespread commercialization, underscoring the urgent need for new materials capable of overcoming these challenges. 

    Ion diffusion in solids, driven by lithium-ion migration through interconnected channels within the crystal structure, is central to the performance of LSICs. The framework of LSICs—comprising mobile lithium ions and immobile lithium-ion-free sublattices—determines the migration pathways and energy distributions\cite{bachman2016inorganic, wang2015design, hull2004superionics}. While some LSICs, like LGPS and Li$_{7}$P$_{3}$S$_{11}$, feature bcc anionic sublattices that facilitate low-energy ion migration, others with non-bcc frameworks, such as garnet and NASICON, also demonstrate high conductivity\cite{jun2022lithium, he2017origin, zhang2019unsupervised}. These findings highlight the limitations of existing structural descriptors in capturing the diverse structural features that contribute to ion transport in LSICs. As such, there is a pressing need for more comprehensive and quantitative methods to understand the structure-function relationship in these materials. While traditional techniques like X-ray Diffraction (XRD) and computational approaches such as graph-based modeling and deep learning have provided valuable insights\cite{zhang2019unsupervised, he2019crystal}, they often overlook the higher-order interactions and topological relationships crucial for accurately predicting ion transport. 
    
    Mathematically, topology encompasses the study of space, connectivity, dimensionality, and transformations. By providing a high level of abstraction, topology serves as an effective tool for analyzing structured data in the physical world, particularly in high-dimensional contexts. However, while topology offers valuable insights, it often oversimplifies geometric information, leading to a loss of structural detail during feature extraction. Persistent homology \cite{edelsbrunner2000topological,zomorodian2005computing}, a burgeoning branch of algebraic topology, presents a promising avenue for reconciling geometry and topology by facilitating a more nuanced understanding of spatial structures in a multiscale topological manner. Persistent homology has found applications in predicting the stability of carbon isomers through the combination of simple linear regression models\cite{xia2015persistent}. Additionally, the introduction of element-specific persistent homology has enabled the preservation of crucial structural information during topological abstraction, particularly beneficial for handling multi-element structures \cite{cang2018integration}. This approach has been successfully employed in predicting the affinity and solubility of molecular proteins in biomedicine\cite{xia2014persistent, wu2018topp}. Furthermore, by restricting its scope of action, persistent homology has been extended to the realm of inorganic crystalline materials exhibiting periodicity. It has proven effective in predicting the formation energies of these materials, showcasing its versatility across different domains\cite{aono1990ionic2}, underscoring its versatility and potential in materials discovery.
    
    Building on these insights, this study introduces a multiscale topological learning (MTL) framework to accelerate the discovery of LSICs. Leveraging persistent homology, the framework extracts multiscale topological features from lithium-ion-only (Li-only) and lithium-ion-free (Li-free) substructures. These substructures are modeled as simplicial complexes to capture higher-order interactions, enabling a more nuanced representation of structural properties. This topological approach preserves critical structural information, offering valuable insights into the spatial organization and functional roles of these substructures in lithium-ion conduction. Next, the present framework introduces two key topological metrics: cycle density ($\rho_{\text{cycles}}$) and minimum connectivity distance ($r_{\text{connected}}$) for quantitative analysis. These metrics quantify the connectivity of Li-only substructures and assess the suitability of Li-free environments for ion diffusion, forming the basis for initial candidate filtering. The resulting materials are further scrutinized with an unsupervised machine learning model, which clusters materials based on similarities in terms of their multiscale topological features. The clustering results indicate that most known LSICs are concentrated within specific clusters, suggesting that other materials in these groups may also exhibit promising ionic conductivity. Finally, a chemical checking process filters out non-LSIC materials, followed by AIMD simulations to validate the remaining candidates. While AIMD simulations are computationally intensive, they are applied exclusively to a small subset of candidates, thereby optimizing resource utilization. This integrated approach not only reduces both computational and experimental costs but also enhances the accuracy of the results, culminating in the identification of 14 novel LSICs and showcasing the efficacy of the proposed framework in accelerating material discovery.


    \section{Results}

    \subsection{Workflow and Conceptual Schematic}  
        
\begin{figure}
      \includegraphics[width=\linewidth]{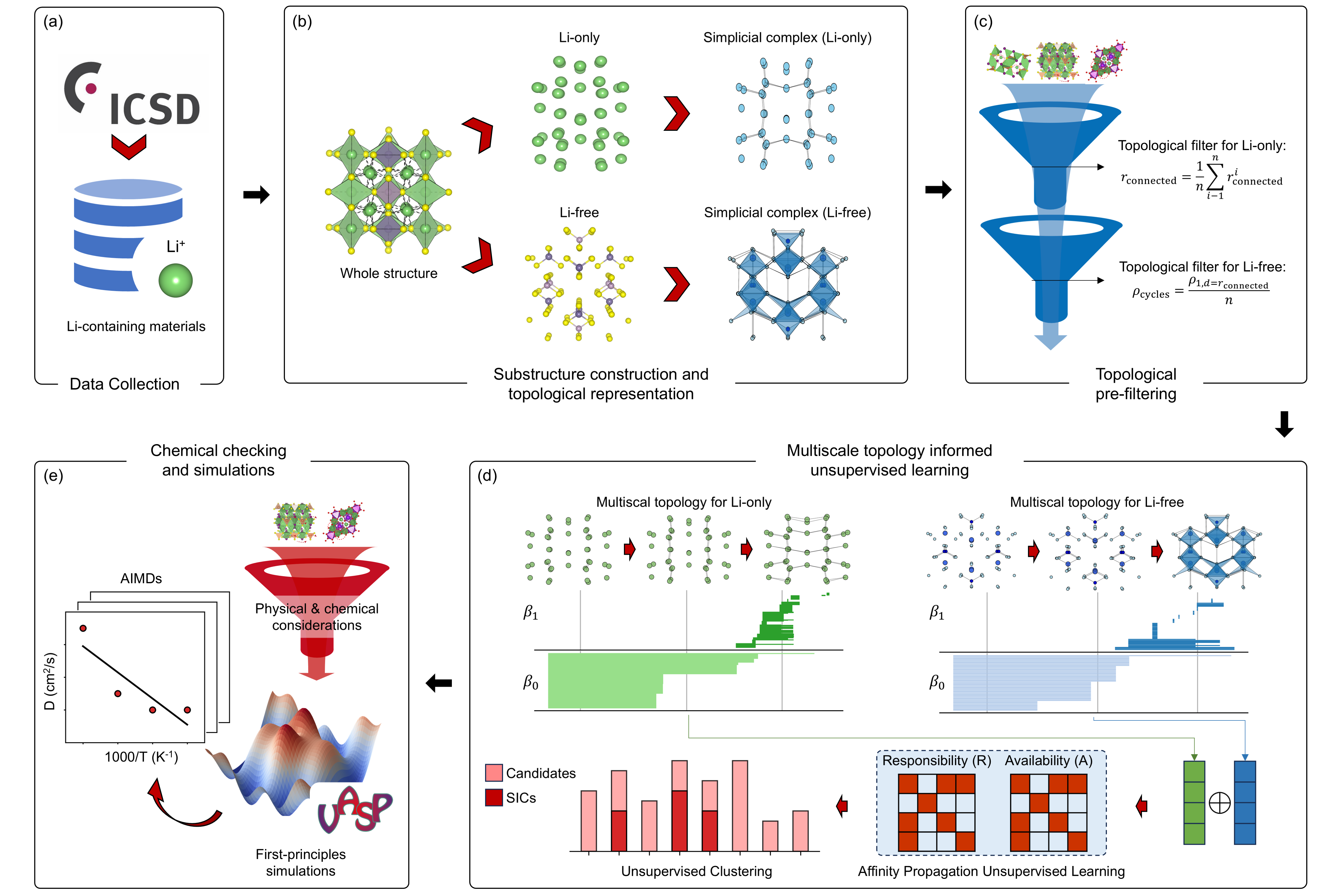}
      \caption{
        Workflow for a multiscale topological learning approach to discovering Lithium superionic conductors.  
        {\bf a} The Data collection phase filters materials containing lithium ions from the ICSD database to identify potential candidates.
        {\bf b} Preliminary study of known LSIC structures, where lithium-only substructures (Li-only) and lithium-free frameworks (Li-free) are modeled as simplicial complexes and analyzed independently.
        {\bf c} Topological representation of Li-only and Li-free substructures using simplicial complexes, capturing high-order interactions and deriving features like connectedness ($r_{connected}$) and cycle density ($\rho_{cycles}$) to narrow the search space.
        {\bf d} Multiscale topological features derived via persistent homology and affinity propagation clustering, grouping materials based on topological similarity to highlight clusters with LSIC candidates.  
        {\bf e} Final physical and chemical validation, including first-principles analysis, to identify the most promising LSIC candidates.
    }
    \label{fig:workflow}
    \end{figure}

        Figure~\ref{fig:workflow} presents the workflow for a multiscale topology approach aimed at discovering Lithium superionic conductors (LSICs). In the initial step (Figure~\ref{fig:workflow}{\bf a}), the data collection phase filters materials containing lithium ions from the ICSD database, identifying promising candidates for analysis. Figure~\ref{fig:workflow}{\bf b} shows the second stage, where a preliminary study of well-known LSIC structures is conducted. Given that ionic conductivity is influenced by both the connectivity of lithium substructures and the stability of the surrounding framework, the Li-only and the Li-free are modeled as independent topological spaces using simplicial complexes and analyzed separately.

        In the next stage, illustrated in Figure~\ref{fig:workflow}{\bf c}, a topological approach is applied to each structure by representing Li-free and Li-only substructures with simplicial complexes. This topological representation captures high-order interactions within the material structure, with each $n$-simplex in the complex representing different types of interactions: 0-simplices (vertices) denote atoms, 1-simplices (edges) capture pairwise atomic interactions, and 2-simplices encode triplet interactions among three atoms. By capturing such high-order interactions, this topological approach provides a deeper structural characterization, essential for understanding ionic conductivity mechanisms in LSICs. Two key features—connectedness ($r_{connected}$) and cycle density ($\rho_{cycles}$)—are derived through this analysis. These features serve as effective filters for narrowing the search space, with $r_{connected}$ encoding information about the Li-only substructure's conductivity and $\rho_{cycles}$ reflecting the stability of the Li-free framework.
        
        In the following stage (Figure~\ref{fig:workflow}{\bf d}), multiscale topological features (persistent homology) are computed through both Li-only and Li-free frameworks, and affinity propagation clustering groups the remaining candidates based on topological feature similarity. This unsupervised clustering reveals internal structural patterns, placing similar materials in proximity within the topological space. Known LSICs tend to cluster within specific groups, highlighting clusters likely to contain additional LSIC candidates. Finally, as shown in Figure~\ref{fig:workflow}{\bf e}, physical and chemical validation, including first-principles-based analysis, is applied to materials within promising clusters. This final evaluation identifies the most viable LSIC candidates, demonstrating the effectiveness of this multiscale topology-based unsupervised learning approach for LSIC discovery.

    \subsection{Topological screening}

\begin{figure}
          \includegraphics[width=\linewidth]{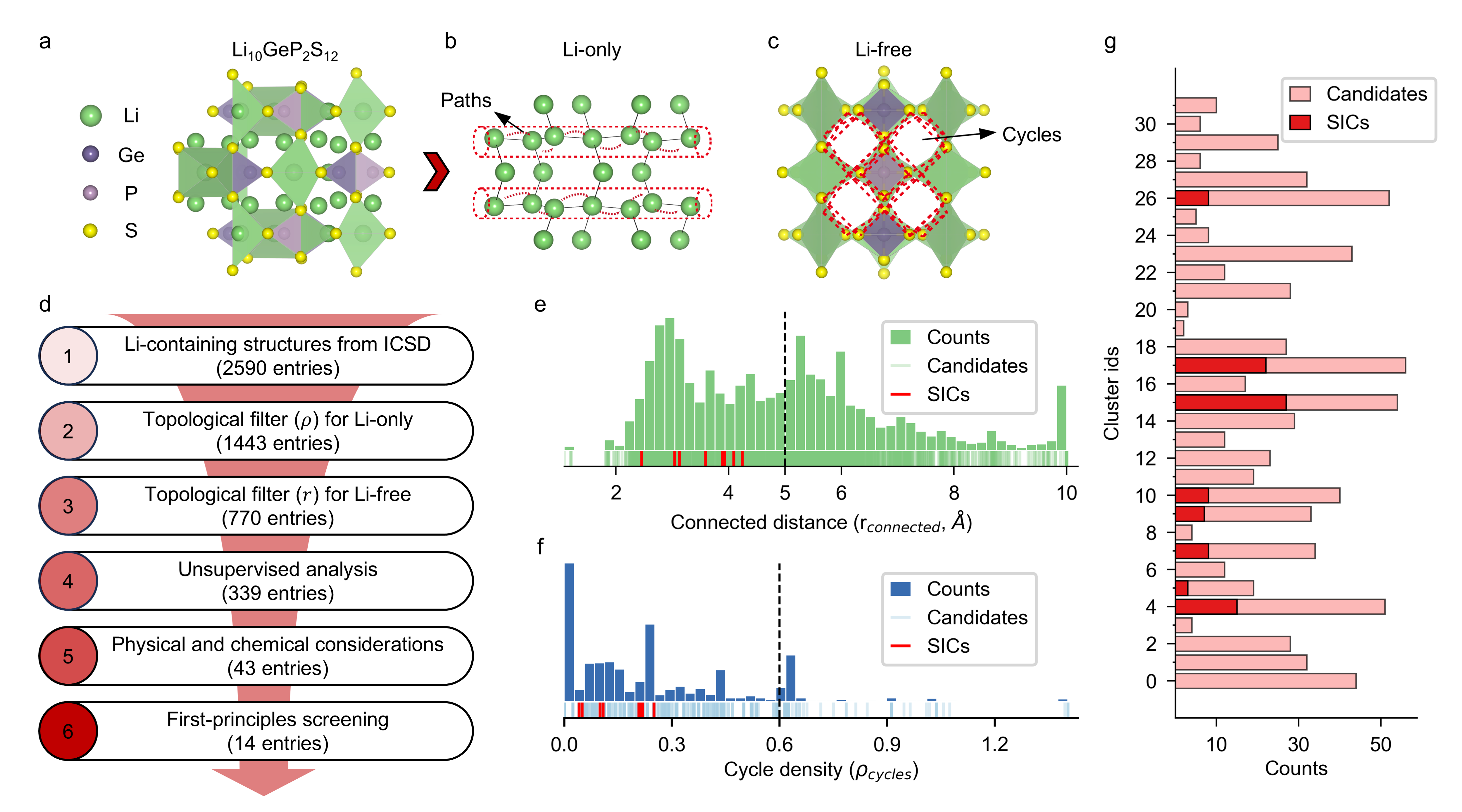}
          \caption{Results of the multiscale topology-driven workflow for LSIC discovery.
            {\bf a} Crystal structure of the solid ionic conductor (LSIC) Li$_{10}$GeP$_2$S$_{12}$, used as an example.  
            {\bf b} The Li-only substructure extracted from the LSIC.
            {\bf c} The Li-free substructure is derived from the same material.  
            {\bf d} Overview of the materials discovery workflow, showing six stages with the corresponding number of materials filtered at each stage.  
            {\bf e} Distribution of the minimum connectivity distances ($r_{\text{connected}}$) for Li-only substructures. The red lines in the rug plot highlight the known LSICs, and the dashed line marks the threshold of $r_{\text{connected}}$ = 5 \AA\ used in the filtering process.  
            {\bf f} Distribution of the pore occupancy index ($\rho_{\text{cycles}}$) for Li-free substructures. The red lines in the rug plot indicate known LSICs, and the dashed line denotes the filtering threshold, $\rho_{\text{cycles}}$ = 0.6.  
            {\bf g} Clustering results from the affinity propagation algorithm. The lighter red bars represent all materials retained after topological pre-filtering, while the darker red bars indicate the known LSICs. The horizontal axis corresponds to the number of structures in each cluster.
          }
        \label{fig:result_rho_r}
        \end{figure}

    Given the limited number of identified LSICs, understanding their internal structural characteristics is essential for advancing materials discovery in this field. In classical LSICs, lithium ions migrate in a cooperative manner characterized by co-diffusion rather than isolated jumping, which is typical of non-super lithium-ion conductors \cite{he2017origin, jalem2013concerted}. This cooperative migration, facilitated by lower energy barriers, indicates that both lithium-lithium interactions and the surrounding framework's structure strongly influence lithium-ion mobility. Additionally, Coulomb interactions among lithium ions affect the migration energy barrier \cite{he2019crystal}. When fractionally or integrally occupied lithium sites are close by (less than 2 \AA\ apart), these interactions produce a continuous lithium-ion probability density within the structure. To fully capture these interactions and effectively identify potential LSICs, it is necessary to analyze both the Li-only and Li-free substructures. Figures~\ref{fig:result_rho_r}{\bf a}-{\bf c} depicts the construction of the Li-only and Li-free substructures from the original material, exemplified by Li$_{10}$GeP$_{2}$S$_{12}$. This process establishes the foundation for identifying LSIC candidates. In the Li-only substructure (Figure~\ref{fig:result_rho_r}{\bf b}), the red channels represent the conductive paths of lithium ions. In the Li-free substructure (Figure~\ref{fig:result_rho_r}{\bf c}), the red cycles illustrate the structural environment surrounding the lithium paths. To streamline the search for suitable LSIC materials among Li-containing compounds, a preliminary filtering process was applied. This filtering process is based on two key topological features—$ r_{\text{connected}} $ and $ \rho_{\text{cycles}} $—that were derived, using a topology-informed approach, for the analysis of Li-only and Li-free frameworks.

    Initially, each Li-only and Li-free substructure was represented as a simplicial complex, an advanced extension of graphs capable of encoding high-order interactions via $ n $-simplices in multidimensional topological space. This complex structure provides a means of analyzing high-order properties that extend beyond pairwise interactions, capturing more intricate atomic configurations through higher-dimensional simplices. By applying algebraic topology techniques, specifically homology and persistent homology, to these simplicial complexes, we extracted topological invariants, known as Betti numbers ($ \beta $), to describe structural features across different dimensions. Here, $ \beta_0 $ denotes the count of independent components, while $ \beta_1 $ represents the number of independent cycles, both of which are essential for assessing material connectivity. Persistent homology was further employed to track changes in these topological invariants across a range of spatial scales. Through a distance-based filtration parameter, the evolution of topological invariants as a function of atomic connectivity was visualized with barcodes, producing unique, scale-dependent topological fingerprints for each structure. An example of topological invariants in the 0, and 1-dimension is shown in Figure~\ref{fig:method}{\bf c}. This approach enables the extraction of key topological and geometric characteristics for both Li-only and Li-free substructures, leading to the development of two essential metrics for filtering materials.

    For the Li-only structure, the metric $ r_{\text{connected}} $ was calculated as the minimum connectivity radius, signifying the critical distance at which all lithium ions in the structure become interconnected. This was determined by taking each lithium-ion within the crystal cell as a center and calculating the connectivity within a spherical region of 10 $\text{\AA}$. The connectivity radius for each lithium-ion was averaged as follows:
    \begin{equation}
        r_{\text{connected}} = \frac{1}{n} \sum_{i=1}^{n} r_{\text{connected}}^i
    \end{equation}
    where $ n $ is the number of lithium sites in the cell. This value provides insight into the minimum connectivity distance required for ion mobility in the Li-only substructure.

    Figure~\ref{fig:result_rho_r}{\bf e} illustrates the distribution of $ r_{\text{connected}} $, a measure of lithium connectivity, calculated for the Li-only substructures of all Li-containing materials in the dataset. The distribution is presented as a histogram with a rug plot shown at the bottom of the figure. Green lines on the rug plot represent the distribution of all materials, while red lines indicate the $ r_{\text{connected}} $ values for known superionic conductors (LSICs), including Li$_7$P$_3$S$_{11}$, NASICON, and LLZO. A detailed list of these LSICs is provided in Table S2. Interestingly, all known LSICs exhibit $ r_{\text{connected}} $ values below 5 \AA\, signifying strong lithium connectivity. This observation highlights a critical characteristic of superionic conductors: the lithium ions are closely paired, ensuring good ionic conductivity. Consequently, a threshold of 5 Å was chosen to screen materials with poor lithium connectivity, effectively narrowing down the dataset from 2,590 to 1,443 materials for further analysis.

    In the Li-free framework, the topological feature $ \rho_{\text{cycles}} $ was derived from the value of $ \beta_1 $ in the topological fingerprint, representing the number of independent ``holes'' or cycles in the structure. These cycles, or voids, within the framework, are essential for facilitating lithium-ion migration. To ensure ionic conductivity, an appropriate number of cycles is required; too many cycles could destabilize the framework, while too few could hinder lithium-ion movement. Here, $ \rho_{\text{cycles}} $ was calculated as:
    \begin{equation}
        \rho_{\text{cycles}} = \frac{\beta_{1, d=r_{\text{connected}}}}{n}
        \end{equation}
    where $ \beta_{1, d=r_{\text{connected}}} $ is the value of $\beta_1$ at $d=r_{\text{connected}}$, and $ n $ denotes the number of lithium sites. This metric captures the balance of voids necessary for ion migration, providing a measure of the Li-free framework's suitability for LSIC functionality.
    
    Figure~\ref{fig:result_rho_r}{\bf f} presents the distribution of $\rho_{\text{cycles}}$, a measure of cycle density, for the Li-free frameworks of the remaining structures after filtering based on $r_{\text{connected}}$. The heights of the histogram bars represent the counts of $\rho_{\text{cycles}}$ values across all Li-free frameworks. At the bottom, a rug plot is shown, where the blue lines indicate the distribution of $\rho_{\text{cycles}}$ for all materials, and the red lines mark the corresponding values for known LSICs. The analysis reveals that effective Li-free frameworks exhibit relatively low cycle density. This finding suggests that a balance is required: the framework must have a sufficient cycle ratio to stabilize the environment surrounding the Li pathways but should not possess excessively high cycle density, which could lead to structural instability or collapse. Based on this observation, a threshold of 0.6 was set for $\rho_{\text{cycles}}$, filtering out unconsolidated Li-free frameworks and refining the selection of candidate materials. The threshold values for both metrics were established based on known LSICs, enabling high-throughput screening of the material database to expedite the identification of potential LSIC candidates.

    \subsection{Multiscale topological clustering}

    In this study, we combined persistent homology, a promising algebraic topology tool, with an unsupervised learning approach to identify potential LSICs among lithium-based materials. Persistent homology offers a robust means of characterizing the structures of both Li-only and Li-free sublattices, providing a comprehensive, multiscale topological fingerprint for each material. The preliminary step used two key topological features derived from the barcodes. The full breadth of features, capturing a more complete spectrum of multiscale topological interactions, was subsequently applied to enhance the clustering process and identify LSIC candidates with greater accuracy.

    To systematically compare materials, we construct fixed-length feature vectors from topology-derived barcodes. For the $\beta_{0}$ of Li-free, since the starting segments of $\beta_{0}$ barcodes are all 0, we extract 7 statistical features from their terminating values: minimum, maximum, mean, sum, standard deviation, median, and $r_{\text{connected}}$. For the $\beta_{1}$ of Li-free, we compute 15 statistics (5 for each of the start, end, and persistence of 1-dimensional barcodes): maximum, minimum, sum, mean, and standard deviation. In total, 22 standardized topological features are generated and used as inputs for an unsupervised learning model to detect potential LSIC candidates. This approach, unlike supervised learning, is well-suited for LSIC discovery, where the scarcity of known LSICs makes supervised training impractical.

    Specifically, the Affinity Propagation (AP) Clustering \cite{frey2007clustering} was employed, which is a graph-based clustering technique that differs from traditional algorithms, such as K-Means, by determining the number of clusters dynamically. The adaptive clustering process enables AP to determine high-quality clusters based on the data's intrinsic structure, avoiding the need for predefined cluster numbers or centroids, enables AP to determine high-quality clusters based on the data's intrinsic structure, avoiding the need for predefined cluster numbers or centroids.
    
    As shown in Figure~\ref{fig:result_rho_r}{\bf g}, the known LSIC materials, represented in dark color, are notably concentrated within a limited number of clusters (8 out of 32), while unclassified materials are shown in lighter shades. The presence of unknown materials within clusters containing known LSICs suggests that these unclassified materials may also exhibit superionic conductivity based on their topological similarity. This multiscale topology-informed unsupervised model enables efficient, label-free identification of LSIC candidates without reliance on predefined hyperparameters or conductivity labels. Ultimately, our approach identified 339 materials clustered alongside known LSICs, providing a refined pool of candidates for further investigation based on their similarity to established LSICs.

    \subsection{Chemical validation and first-principles verification}

    \begin{figure}
        \centering
        \includegraphics[width=\linewidth]{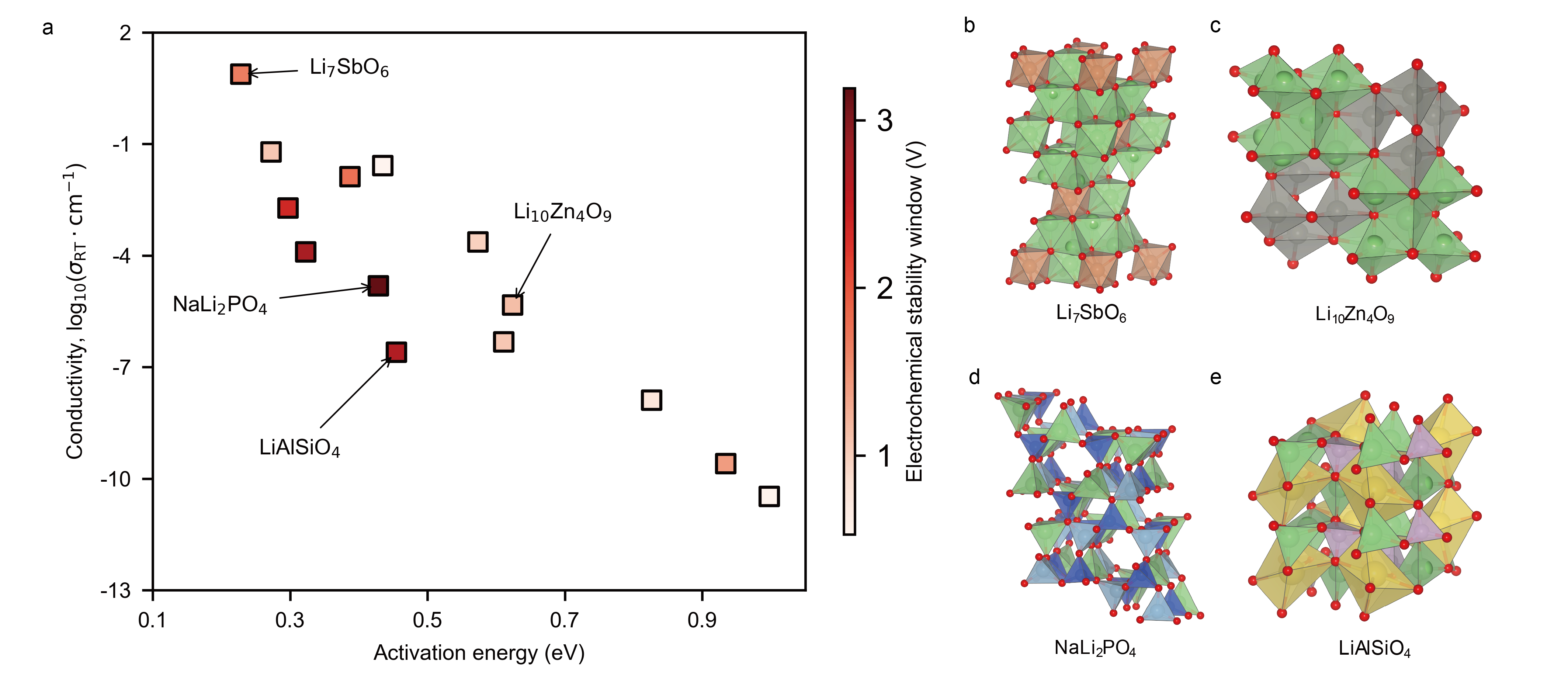}
        \caption{ 
            {\bf a} The ionic conductivity as a function of the lithium-ion diffusion activation barrier for 14 potential LSICs, was calculated using AIMD at room temperature (300 K). The color bar indicates the electrochemical stability window of these materials.  
            {\bf b}, {\bf c}, {\bf d}, and {\bf e} demonstrate the structures of Li$_7$SbO$_6$, Li$_{10}$Zn$_4$O$_9$, LiAlSiO$_4$ and NaLi$_2$PO$_4$.
        }
      \label{fig:conductivity}
      \end{figure}

    To further validate the LSIC candidates filtered through the unsupervised learning model, a rigorous chemical screening process was applied to ensure both structural and chemical suitability for practical applications.  Several criteria were established for this stage of validation: (1) compounds with two or fewer elements were excluded; (2) materials containing more than 500 atoms were removed due to computational limitations and challenges in experimental validation; (3) compounds with radioactive elements or water molecules were eliminated; (4) alloys were excluded; (5) compounds with elements in abnormal oxidation states, which could compromise stability, were removed; (6) specific classes of compounds, such as all Li-X-O ternary systems where X is S, I, Si, C, P, Al, Ge, Se, B, or Cl, and Li-P-S systems, were excluded; (7) compounds containing transition metals like Fe, Mn, Ni, Ti, Mo, V, Co, and others, or oxide compounds with N, Re, Ho, Hf, Ru, Eu, Lu, were omitted; and (8) compounds in which lithium shared atomic sites with other elements were excluded to avoid hindrance of lithium-ion diffusion channels. A total of 339 alternative materials were subjected to this screening process, as detailed in Table S3, ultimately narrowing the pool to 44 candidates (Table S4). 

    Following the chemical screening phase, AIMD simulations were employed to evaluate the ionic conductivity, lithium-ion diffusion activation barriers, and electrochemical stability of the 45 selected materials. These simulations were conducted at elevated temperatures (800 K, 1000 K, 1200 K, and 1400 K) to accurately capture lithium-ion diffusion behavior and calculate activation barriers, as detailed in Table S5. By integrating these results with electrochemical stability window (ESW) calculations, the analysis provided a comprehensive assessment of the structural and dynamic properties of the candidates.

    To balance conductivity and stability, thresholds were established based on experimental and computational guidelines. Lithium-ion diffusion activation barriers were constrained between 0.1 and 1.0 eV, ensuring the exclusion of materials with impractically low barriers, which may indicate structural instability, while allowing for sufficient ionic mobility. Candidates who pass the threshold of activation barriers are shown in Table S6. Similarly, an ESW threshold of 0.5 V was applied to ensure chemical stability under slightly reducing conditions, such as those encountered during cycling, as detailed in Table S6. These thresholds prioritize materials that achieve an optimal balance between high ionic conductivity, structural stability, and compatibility with lithium-metal anodes or other battery components.

    From this comprehensive analysis, 14 materials were identified that satisfied the desired criteria. Figure~\ref{fig:conductivity}{\bf a} illustrates the ionic conductivity as a function of the diffusion activation barriers for these final candidates, many of which demonstrate excellent ionic conductivity in the order of $10^{-2}$ S/cm at room temperature (300 K). Detailed results for these candidates, including Arrhenius plots of lithium-ion diffusion coefficients, structural representations, and isosurfaces of lithium-ion probability densities obtained from AIMD simulations, are provided in Figures S1-S14 and Table 1.

    This multi-stage validation process highlights the importance of integrating chemical screening with structural and dynamic assessments to identify high-potential LSICs. Notably, several materials identified in this study—such as Li$_7$SbO$_6$, Li$_{10}$Zn$_4$O$_9$, and LiAlSiO$_4$—have been independently validated experimentally or patented, further substantiating the approach's predictive power. For instance, Li$_7$SbO$_6$ (Figure~\ref{fig:conductivity}{\bf b}) demonstrated excellent rate performance, high cycling stability, and outstanding Coulombic efficiency, making it well-suited for high-rate lithium battery applications\cite{Li7SbO6}. The formation of the Li$_10$Zn$_4$O$_9$ (Figure~\ref{fig:conductivity}{\bf c}) nanophase is considered one of the primary factors contributing to the high conductivity in glassy lithium-ion conductors, indicating that this nanophase plays a crucial role in enhancing overall ionic conductivity\cite{Li10Zn4O9}. LiAlSiO$_4$ (Figure~\ref{fig:conductivity}{\bf d}), with its high transparency, excellent ionic conductivity, and cost-effectiveness, demonstrates significant potential as an electrolyte in high-performance all-solid-state electrochromic devices\cite{LiAlSiO4}. Yet, in experiments, the ionic conductivity of the LiAlSiO$_4$ thin film was approximately 2.7 × 10$^{-5}$ mS/cm. Additionally, it was observed that the structural type of NaLi$_2$PO$_4$ (Figure~\ref{fig:conductivity}{\bf e}) is similar to that of Li$_3$PO$_4$. As a well-known solid-state electrolyte, the framework structure of Li$_3$PO$_4$ indicates that NaLi$_2$PO$_4$ has potential as a solid-state electrolyte material\cite{Li3PO4first, Li3PO4second}.

    These findings validate the proposed model's efficacy in identifying promising LSICs and emphasize its potential to accelerate the discovery of advanced materials for next-generation lithium-ion batteries. Moreover, the identified candidates that have yet to be experimentally tested present exciting opportunities for future research, demonstrating the robustness and scalability of the methodology.

    \begin{table*}[t]
        \centering
        \caption{Potential LSICs filtered through AIMD simulations, including ICSD IDs and corresponding calculated properties.}
        \label{tbl:filtered_SICs}
        \resizebox{1\textwidth}{!}{
        \begin{tabular}{cccccc}
        \hline
        ICSD-IDs & Compositions	& Structure Type & \makecell[c]{Activation \\Barrier (eV)}	&\makecell[c]{Ionic Conductivity (mS/cm)}	& \makecell[c]{Electrochemical \\ Stability Window (V)} \\ 
        \hline
        9987 & Li$_6$Ga$_2$(BO$_3$)$_4$ & Li$_3$AlB$_2$O$_6$ & 0.826 & 1.291e-8 & 1.654 \\
        \hline
        15631 & Li$_7$SbO$_6$\cite{Li7SbO6} &   & 0.228 & 7.634 & 1.137 \\
        \hline
        23634 & Li$_{10}$Zn$_4$O$_9$\cite{Li10Zn4O9} &   & 0.624 & 4.694e-6 & 1.405 \\
        \hline
        35250 & K$_2$Li$_{14}$Pb$_3$O$_{14}$ & K$_2$Li$_{14}$Pb$_3$O$_4$  & 0.998 & 3.291e-11 & 0.553 \\
        \hline
        40245 & Li$_3$BiO$_3$ &  & 0.573 & 2.32e-4 & 0.977 \\
        \hline
        59640 & Li$_4$Zn(PO$_4$)$_2$ & Li$_4$O$_8$P$_2$Zn & 0.387 & 1.32e-2 & 1.772 \\
        \hline
        69967 & NaLi$_2$PO$_4$ & Li$_3$PO$_4$\cite{Li3PO4first, Li3PO4second} & 0.429 & 1.52e-5 & 3.19 \\
        \hline
        71035 & KLi$_6$BiO$_6$ & KLi$_6$IrO$_6$ & 0.272 & 6.16e-2 & 1.064 \\
        \hline
        72840 & Li$_6$KBiO$_6$ &  & 0.611 & 4.792e-7 & 1.063 \\
        \hline
        74864 & CsKNa$_2$Li$_8$(Li(SiO$_4$))$_4$ & CsKNa$_2$Li$_8$(LiSiO$_4$)$_4$ & 0.296 & 1.89e-3 & 2.382 \\
        \hline
        78819 & Li$_{10}$N$_3$Br &  & 0.435 & 2.60e-2 & 0.530 \\
        \hline
        92708 & LiAlSiO$_4$\cite{LiAlSiO4} & LiGaSiO$_4$ & 0.455 & 2.504e-7 & 2.667 \\
        \hline
        95972 & Li$_2$MgSiO$_4$ & Li$_2$ZnSiO$_4$ & 0.323 & 1.25e-4 & 2.739 \\
        \hline
        262642 & In$_2$Li$_2$SiS$_6$ & Cd$_4$GeS$_6$ & 0.934 & 2.546e-10 & 0.755 \\
        \hline
        \end{tabular}}
        \end{table*}

    \subsection{Discussions}
    This study highlights the effectiveness of a multiscale topology analysis approach, integrated with unsupervised learning, for quantitatively characterizing lithium-ion diffusion channels and their surrounding frameworks within crystal structures. The workflow and filtered structures at each stage are summarized in Figure~\ref{fig:result_rho_r}{\bf d}. A high-throughput topological analysis of lithium-containing compounds provided quantitative insights into their crystal structures and significantly narrowed the pool of potential LSIC candidates. 

    The initial phase of the strategy reduces the search space by analyzing two critical factors: cycle density ($\rho_{\text{cycles}}$) for lithium-free substructures (Li-free) and minimum connectivity distance ($r_{\text{connected}}$) for lithium-only substructures (Li-only). This dual-filtering approach ensures the retention of structures meeting the essential criteria for lithium-ion diffusion and stable frameworks. Specifically, all identified LSICs exhibit Li-free sublattices with $\rho_{\text{cycles}}$ below 0.6, ensuring a balanced cycle density conducive to stability, and Li-only sublattices with $r_{\text{connected}}$ below 5 \AA, enabling efficient ionic conduction.

    Following this filtering step, the strategy leverages multiscale topology-based features to further refine the candidate pool. These features capture both the global structural characteristics, using a multiscale filtration process, and the inherent properties of the structures, encompassing ionic transition pathways and their environmental frameworks. By comparing these refined candidates with known LSIC structures through affinity propagation clustering, the method effectively identifies potential LSICs. This unsupervised learning step highlights materials structurally similar to known LSICs while uncovering novel, previously unstudied materials. This approach successfully identified all known LSIC structures and revealed 45 additional potential LSIC candidates.

    The proposed strategy demonstrates a highly efficient method for LSIC discovery by integrating advanced mathematical frameworks and machine learning techniques. The initial focus on two key topological features, combined with a comprehensive multiscale topological analysis, efficiently narrows a vast dataset while maintaining high predictive accuracy. Moreover, the approach's generalizable framework can be extended to other materials of interest, offering a scalable and innovative pathway for materials discovery.

    To validate these refined candidates, more precise AIMD simulations were conducted to assess their ionic conductivity, lithium diffusion activation barriers, and electrochemical stability. Among the candidates, 14 materials met stringent criteria, including a lithium-ion diffusion activation barrier below 1.0 eV and an electrochemical stability window greater than 0.5 V. Several of these materials have been experimentally validated as excellent LSICs, further confirming the model's predictive capability. The remaining candidates offer promising avenues for future experimental evaluation. Overall, this robust and efficient workflow ensures the discovery of materials with desired properties, even when only limited verified knowledge is available.

    This work demonstrates the potential of combining advanced topological methods with unsupervised learning for efficient material discovery. The proposed methodology is not limited to LSICs. It can be adapted to discover other materials with desired properties, providing a versatile and generalizable strategy for addressing complex challenges in materials science.

    \section{Methods}\label{Sec:Method}
     
     \subsection{Multiscale topology data analysis}

\begin{figure}
            \includegraphics[width=\linewidth]{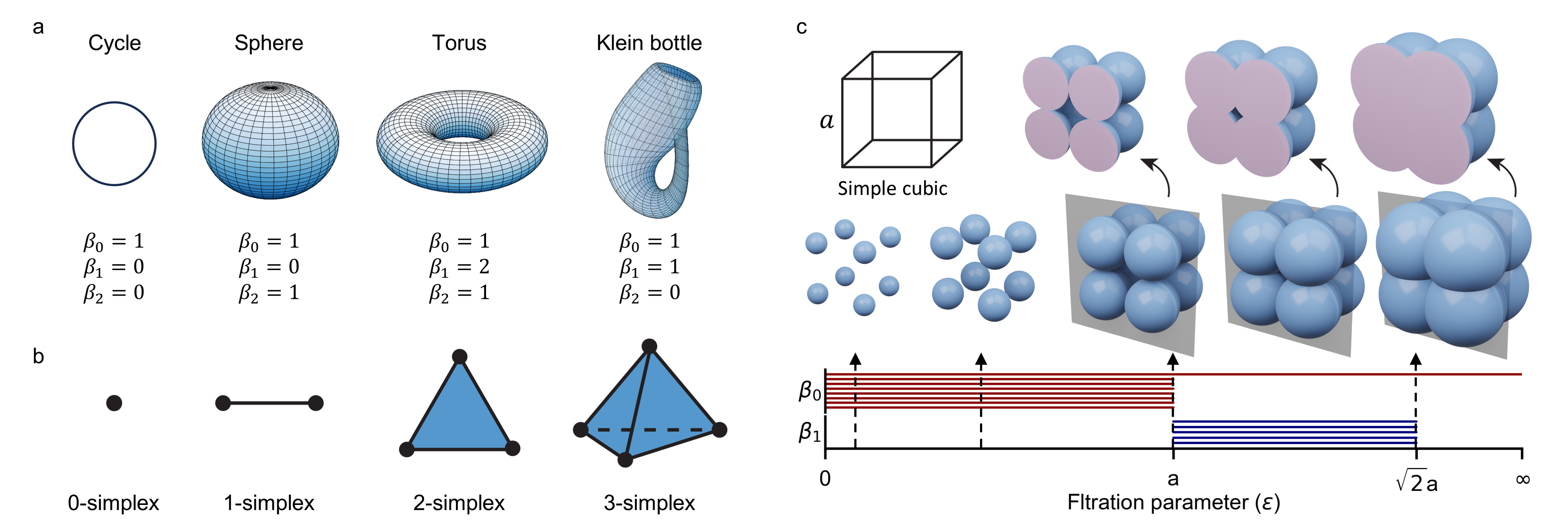}
            \caption{
                {\bf a} Examples of topological spaces and their Betti numbers. A cycle has $\beta_0 = 1$, $\beta_1 = 1$, $\beta_2 = 0$; a sphere has $\beta_0 = 1$, $\beta_1 = 0$, $\beta_2 = 1$; a torus has $\beta_0 = 1$, $\beta_1 = 2$, $\beta_2 = 1$; and a Klein bottle exhibits non-trivial Betti numbers with $\beta_0 = 1$, $\beta_1 = 1$, $\beta_2 = 0$.
                {\bf b} Building blocks of simplicial complexes, represented by simplices of increasing dimensions: vertices (0-simplices), edges (1-simplices), triangles (2-simplices), and tetrahedra (3-simplices).
                {\bf c} Workflow of persistent homology illustrated using a Vietoris-Rips complex. A simple cubic structure is analyzed by progressively increasing a filtration parameter $d$, which expands balls around each vertex. As $d$ grows, topological features such as connected components ($\beta_0$) and loops ($\beta_1$) emerge and persist. The persistence of cycles in each phase of the cubic structure is visualized through barcodes corresponding to $\beta_1$.
            }
          \label{fig:method}
          \end{figure}

        \paragraph{Simplicial complex representation}
        In this work, both Li-free and Li-only structures are modeled using simplicial complexes, which extend graphs to higher dimensions, providing richer structural and topological insights. A simplex, the building block of a simplicial complex, generalizes geometric shapes like points (0-simplices), line segments (1-simplices), triangles (2-simplices), and tetrahedra (3-simplices) to arbitrary dimensions, as shown in Figure~\ref{fig:method}{\bf b}. For material representation, atoms are treated as 0-simplices (vertices), and atomic interactions are captured by higher-dimensional simplices, reflecting structural hierarchy and connectivity. A $k$-simplex, defined as $\sigma^{k} = \Big\{ v \,|\, v = \sum_{i=0}^{k} \lambda_{i} v_{i}, \sum_{i=0}^{k} \lambda_{i} = 1, 0 \leq \lambda_{i} \leq 1 \Big\}$, is the convex hull of $k+1$ affinely independent points. A simplicial complex $K$ is a collection of simplices satisfying: (1) Every face of a simplex in $K$ is also in $K$; (2) The intersection of any two simplices is either empty or a common face.

        \paragraph{Homology and persistent homology}
        Homology provides an algebraic framework to analyze simplicial complexes, revealing topological features such as connectedness, holes, and voids across dimensions. Central to this framework are chains, chain groups, chain complexes, and boundary operators. A $k$-chain is a formal sum of $k$-simplices with coefficients in a chosen field (e.g., $\mathbb{Z}_2$), and the collection of all $k$-chains forms the $k$-chain group $C_k$. The boundary operator $\partial_k$ maps $k$-chains to $(k-1)$-chains:  
        \begin{equation}
        \partial_k \sigma^{k} = \sum_{i=0}^{k} (-1)^i [v_0, \ldots, \hat{v}_i, \ldots, v_k],
        \end{equation}
        where $\hat{v}_i$ omits the $i$-th vertex. This operator defines cycles ($\text{Ker}(\partial_k)$: chains with no boundary) and boundaries ($\text{Im}(\partial_{k+1})$: chains that are boundaries of higher-dimensional simplices). These relationships form a chain complex:  
        \begin{equation}
        \cdots \xrightarrow{\partial_{k+1}} C_k \xrightarrow{\partial_k} C_{k-1} \xrightarrow{\partial_{k-1}} \cdots \xrightarrow{\partial_1} C_0 \xrightarrow{\partial_0} 0,
        \end{equation}
        where $\partial_{k-1} \circ \partial_k = 0$. The $k$-th homology group $H_k$ is defined as: $H_k = \text{Ker}(\partial_k) / \text{Im}(\partial_{k+1})$, and measures $k$-dimensional holes in the simplicial complex. The Betti numbers $\beta_k = \text{rank}(H_k)$ quantify the number of independent $k$-dimensional features, such as connected components ($\beta_0$), tunnels ($\beta_1$), and cavities ($\beta_2$). Figure~\ref{fig:method}{\bf a} shows the examples of topological spaces and their Betti numbers, a cycle has $\beta_0 = 1, \beta_1 = 0, \beta_2 = 0$, while more complex shapes such as the torus and Klein bottle have non-trivial higher-dimensional Betti numbers.

        Persistent homology extends homology to multiscale analysis, capturing the persistence of topological features as a parameter (e.g., a scale parameter $\epsilon$) varies \cite{edelsbrunner2000topological,zomorodian2005computing}. This is achieved through filtration, a sequence of nested simplicial complexes $\{K_i\}$ such that $K_0 \subseteq K_1 \subseteq \dots \subseteq K_n$. This work uses the Vietoris-Rips filtration, where simplices are added based on a distance threshold $\epsilon$. Persistent homology tracks the evolution of homological features through filtration steps:  
        \begin{equation}
        \emptyset = H(K^0) \rightarrow H(K^1) \rightarrow \dots \rightarrow H(K^n) = H(K).
        \end{equation}

        The $p$-persistent $k$-th homology group describes features persisting across filtration steps $i$ to $i+p$: $H_{k}^{i,p} = Z_{k}^{i} / (B_{k}^{i+p} \cap Z_k^{i})$,where $Z_k^{i}$ and $B_k^{i+p}$ are the cycles and boundaries at steps $i$ and $i+p$, respectively. Persistent homology is often visualized using barcodes, where each bar represents a topological feature's birth and death as $\epsilon$ increases. Figure~\ref{fig:method}{\bf c} illustrates a simple cubic at varying thresholds $\epsilon$ and their corresponding persistent patterns.

     \subsection{Clustering}
        The Affinity Propagation (AP) algorithm is a clustering technique designed to identify a set of exemplars among data points and assign each point to its nearest exemplar, forming distinct clusters \cite{frey2007clustering}. Unlike traditional clustering methods like K-Means, AP does not require pre-specifying the number of clusters. Instead, it dynamically determines the clusters based on the similarities among data points.

        The algorithm begins by calculating the pairwise similarity between data points. For data points $ \mathbf{x}_i $ and $ \mathbf{x}_k $, the similarity is defined as $ s(i, k) = -\| \mathbf{x}_i - \mathbf{x}_k \|^2 $, which measures how well $ \mathbf{x}_k $ can serve as the exemplar for $ \mathbf{x}_i $. Two key matrices, the responsibility matrix ($ \mathbf{R} $) and the availability matrix ($ \mathbf{A} $), are then iteratively updated to identify exemplars.
        These updates continue until the algorithm converges, producing exemplars that maximize cluster similarity. Each data point is assigned to the cluster corresponding to its most suitable exemplar, defined by the combination of responsibility and availability scores. This iterative process ensures robust cluster formation without requiring predefined parameters like the number of clusters.
        
        For this study, the implementation of AP from the scikit-learn library was employed \cite{scikit-learn}. This method's ability to dynamically identify cluster centers makes it particularly suitable for analyzing the complex, high-dimensional feature space generated by the multiscale topological method. It facilitated the identification of clusters representing structurally and chemically similar materials, enabling effective material categorization and candidate screening.

     \subsection{First-principles simulation}
     In this work, all Density Functional Theory (DFT) calculations were performed using the Vienna Ab Initio Simulation Package (VASP), utilizing the Projector Augmented Wave (PAW) method in conjunction with the Perdew-Burke-Ernzerhof (PBE) exchange-correlation functional \cite{kresse1996efficiency, perdew1996rationale}. The plane wave basis set employed a cutoff energy of 520 eV to ensure computational accuracy and efficiency. For structural optimization, k-point meshes centered on the $ \Gamma $-point were generated with a minimum spacing of 0.4 \AA\ between k-points. A finer k-point spacing of 0.25 \AA\ was used for accurate energy calculations.

     Ab Initio Molecular Dynamics (AIMD) simulations were conducted to assess lithium-ion diffusion. The systems were first relaxed, then heated to 1200 K over 10 ps, followed by equilibration at 800 K, 1000 K, 1200 K, and 1400 K for 20 ps, excluding the initial 2 ps of each trajectory. A time step of 2 fs was used for the AIMD simulations with $ \Gamma $-point k-point sampling.

    Ionic diffusivity ($ D $) was calculated using the mean square displacement (MSD) formula:
    \begin{equation}
        D = \frac{1}{2dN \Delta t} \sum_{i=1}^{N} \left\langle \left[ \mathbf{r}_i(t + \Delta t) - \mathbf{r}_i(t) \right]^2 \right\rangle_t
    \end{equation}
    where $ d $ is the dimensionality of diffusion, $N$ is the number of ions, and $ \mathbf{r}_i(t) $ is the displacement of the $ i $-th ion.

    The ionic conductivity ($ \sigma $) was then derived using the Nernst-Einstein relation:
    \begin{equation}
        \sigma = \frac{n q^2}{k_B T}D
    \end{equation}
    where $ n $ is the ion density, $ q $ is the ion charge, $ k_B $ is the Boltzmann constant, and $ T $ is the temperature. These calculations provided key insights into the ionic transport properties of the materials, including diffusion coefficients, lithium-ion diffusion activation barriers, and electrochemical stability windows.

    \section*{Data availability}
        The dataset used in this study is from the ICSD database, and all data can be downloaded from the official ICSD website. Additionally, we have provided a list of ICSD numbers for the data at each filtering step on \url{https://github.com/PKUsam2023/MTUL-LSIC/tree/main/filter_data}.

    \section*{Code availability}
        The related codes have been released as an open resource in the Github repository: \url{https://github.com/PKUsam2023/MTUL-LSIC/tree/main}.

    \section*{Supporting Information}
        The Supporting Information is available on the website at \url{xxxxxx}

    \section*{Acknowledgment}
        This work is financially supported by the Shenzhen Science and Technology Research Grant (No. ZDSYS201707281026184), the Soft Science Research Project of Guangdong Province (No. 2017B030301013), and The Major Science and Technology Infrastructure Project of Material Genome Big-science Facilities Platform supported by Municipal Development and Reform Commission of Shenzhen. The work of Chen and Wei was supported in partial by NIH grants R01GM126189, R01AI164266, and R35GM148196, NSF grants DMS-2052983 MSU Research Foundation, and Bristol-Myers Squibb 65109.

    \section*{Conflict of Interest}
        The authors declare no conflict of interest.

    
\end{document}